# Control of catalytic efficiency by a co-evolving network of catalytic and non-catalytic residues

~~A co-evolutionary barrier constrains active site variation~~

~~in LAGLIDADG homing endonucleases~~

Thomas A. McMurrough, Russell J. Dickson, Stephanie M. F. Thibert, Gregory B. Gloor* and David R. Edgell*

Department of Biochemistry

Schulich School of Medicine and Dentistry

Western University

London, ON

N6A 5C1

Canada

* Corresponding authors: dedgell@uwo.ca, ggloor@uwo.ca

**The active sites of enzymes consist of residues necessary for catalysis, and structurally important non-catalytic residues that together maintain the architecture and function of the active site. Examples of evolutionary interactions between active site residue have been difficult to define and experimentally validate due to a general intolerance to substitution[1]. Here, using computational methods to predict co-evolving residues[2], we identify a network of positions consisting of two catalytic metal-binding residues and two adjacent non-catalytic residues in LAGLIDADG family homing endonucleases (LHEs)[3]. Distinct combinations of residues in the network map to clades of LHE sub-families, with a striking distribution of the metal-binding Asp (D) and Glu (E) residues. Mutation of these four positions in two LHEs, I-LtrI and I-OnuI[4], indicate that combinations of residues tolerated are specific to each enzyme. Kinetic analyses under single-turnover conditions[5,6] indicated a substantial defect in either the catalytic rate ($k_{cat}*$) or the Michaelis constant ($K_M*$) for I-LtrI variants with networks predicted to be suboptimal. Enzymatic activity could be restored to near wild-type levels with a single compensatory mutation that recreated a network of optimally predicted residues. Our results demonstrate that LHE activity is constrained by an evolutionary barrier of residues with strong context-dependent effects. Creation of optimal co-evolving active site networks is therefore an important consideration in engineering of LHEs and other enzymes.**

Networks of co-evolving amino acids are thought to be important for protein structural integrity, domain and allosteric interactions, yet are rarely considered to directly contribute to enzymatic active sites[1]. This may be because active site residues are highly conserved, and since they do not vary, they cannot co-vary and co-evolution with surrounding residues cannot be inferred[7]. Here, we have addressed the question of co-evolution between active site and surrounding non-catalytic residues using LHEs, site-specific DNA endonucleases that are typically encoded within self-splicing introns and inteins[8]. LHEs show extreme sequence variation[9,10] and function as homodimers or as single-chain monomers composed of two LAGLIDADG domains that evolved by gene duplication or gene fusion events. The active site is composed of two parallel α-helices with acidic metal-binding residues at the bottom of each helix positioned in close proximity to the DNA substrate[11,12]. Integrity of the helix interface is crucial for function[13,14], and previous attempts to optimize this interface have met with variable outcomes[15,16]. We noted that the occurrence of two different residues, Asp (D) and Glu (E), at the catalytic metal-binding site of the single-chain LHEs is an unusual feature for a protein family, and might be influencing enzyme activity. In I-LtrI[4], the metal-binding residues corresponded to positions E29 and E184 (I-LtrI numbering is adopted throughout the manuscript). The evolutionary distribution of these residues was examined with an alignment of 178 LHE protein sequences generated by a structure-based method and curated to remove all contaminating maturase sequences[17] (Supplementary Fig. 1). Unexpectedly, a clear clade preference was found when the distribution of the catalytic residues was mapped onto a maximum likelihood phylogenetic tree derived from the full alignment by PhyML (Fig. 1a and Supplementary Fig. 2). We considered the possibility that the clade preference reflected an evolutionary and functional interaction between the metal-binding and other residues in the α helices. Examination of the alignments by sequence covariation analysis[2] showed that LAGLIDADG domain α-helices 1 and 2 were the most strongly covarying segments of the protein family by $Zpx$ scores, with the residue positions immediately preceding the metal-binding residues (A28 and G183 of I-LtrI) exhibiting the strongest covariation signal with each other (Fig. 1b and c and Supplementary Table 1).

Surprisingly, the positions corresponding to the metal-binding sites were also identified in this analysis.

Based on these observations we examined the initial hypothesis that the clade-specific distribution of the metal-binding residues was influenced by a covarying network composed of these four positions. The first test of this hypothesis examined the distribution and frequency of all observed quartet species in the phylogenetic tree (Fig.1a and 1d, and Supplementary Table 2). The results indicate that there was a strong association between the phylogenetic position of the metal binding residue and the adjacent residues, and between the identity of the metal-binding residues and the identity of the residues permitted at the adjacent positions. However, the achievement of statistical rigor is difficult from such analyses because of the small number of non-independent sequences sampled.

An unbiased screen was performed to determine if the observed phylogenetic relationship could be recapitulated in an experimental system that selects for active enzyme variants. Positions 28 and 183 were each randomized in the context of the I-LtrI backbone to all 20 amino acids, while the metal-binding residues at positions 29 and 184 were each held at either D or E to give a complexity of 1600 possible I-LtrI variants (Fig. 2a). This library was screened for enzymatic activity in a liquid culture assay where expression of an active endonuclease (carried on pEndo-Ltr) would cleave an I-LtrI recognition site on a plasmid encoding a bacteriostatic DNA gyrase toxin gene, thereby eliminating the toxin-carrying plasmid (pTox-Ltr)[18]. The rate of destruction of pTox-Ltr by the I-LtrI variants correlates with the ability of the enzyme to bind to and cleave the recognition site. Loss of pTox-Ltr is reflected as an increase in relative abundance for active I-LtrI variants over weak or inactive variants (Fig. 2b).

We performed 7 independent transformations of the pEndo-Ltr library into pTox-Ltr competent cells, harvested total plasmid DNA from selective and non-selective conditions after 16 hours outgrowth, and examined the mutated positions by Illumina paired-end sequencing. Because the strategy captured all four positions in paired-end sequencing reads, it was possible to determine the enrichment of all 1600 possible quartet combinations in the selected condition versus the unselected condition[19] (Supplementary Table 3 and Supplementary Fig. 3). As shown in Figure 2b, I-LtrI variants with small side chain residues at positions 28 and 183 were strikingly enriched relative to other combinations, indicating that these positions were not randomly assorting. In particular, combinations of A and G were favoured, possibly indicating that steric restrictions at the base of the LAGLIDADG domain α-helices comprising and surrounding the active site may influence residue identity. Residue combinations at positions 28 and 183 that occurred at low frequency in the LHE alignment, or that were not observed, were also enriched (V_G, T_G, S_G), confirming that the frequency of residues in the current LHE alignment reflects an ascertainment bias. Variants with D_E or D_D combinations at positions 29 and 184 within the context of the I-LtrI backbone were underrepresented relative to I-LtrI variants with E_D or E_E combinations. These observations strongly parallel the clade preference of the metal-binding residues on the LHE phylogeny (Fig. 1a), supporting the hypothesis that a covarying network of residues influences the amino acid composition of these positions. We conclude that the distribution of metal-binding residues within related LHEs is likely constrained by these context-dependent interactions.

To investigate the basis for the low abundance of D_D and D_E variants in the liquid competition experiments, we tested survival of individual variants in a bacterial two-plasmid survival assay[18] on solid media and found that I-LtrI variants with sub-optimal combinations at positions 28_183 and 29_184 survived poorly, or not at all (Fig. 3a, left panel and Supplementary Table 4). Similar findings were found when the same mutations were made in I-OnuI (Fig. 3a, right panel and Supplementary Table 5), a LHE enzyme with ~30% sequence identity to I-LtrI. Variants with G_A or A_G at positions 28_183 in combination with D_D and D_E at the metal-binding positions survived, but had small colony sizes relative to E_D or E_E variants (Supplementary Fig. 4). This phenotype was supported by liquid growth experiments that measured the time individual variants took to reach mid-log phase ($A_{600}$=0.35) (Fig. 3b and Supplementary Fig. 5). Notably, D_D or D_E metal-binding residue variants all had longer times to reach mid-log phase than E_E or E_D variants with the same residues at positions 28_183. Collectively, these data show that the interaction between the metal-binding and adjacent residues is generalizable to other LHE sequence backbones, and that the co-evolving residues modulate survival and growth rate.

We ruled out the possibility that low enrichment and low survival in the competition and plate experiments was due to structural defects by purifying select variants and determining their thermal stability. Differential scanning calorimetry analyses showed that the melting temperatures and enthalpy of denaturation were not obviously different when either the metal-binding or adjacent residues were mutated in I-LtrI or I-OnuI (Supplementary Table 5 and Table 6). To test the hypothesis that the covarying network modulates the catalytic efficiency of LHE enzymes, the psuedo-Michaelis-Menten parameters[5,6] $k_{cat}$* and $K_M$* were determined from single-turnover reaction conditions (Fig. 3c and Supplementary Fig. 6, Fig. 7 and Fig. 8). For example, a single change at position 183 (G183A) to create the suboptimal A_A:E_E network resulted in a ~65-fold decrease in $k_{cat}$*, whereas a single change at position 28 (A28G) to create G_G:E_E generated an enzyme with a ~3.5-fold increase in $K_M$* over the wild-type enzyme (Fig. 3c). Interestingly, the G_G combination is infrequently found in the LHE alignment (Fig. 1d) and exhibits slower growth compared to A_G variants (Fig. 2b and 3a), suggesting a penalty for highly efficient enzymes in a biological context. Plotting catalytic efficiency ($k_{cat}$*/$K_M$*) versus time to mid-log phase in liquid culture for I-LtrI network variants revealed a striking correlation (Fig. 3d and Supplementary Fig. 9), suggesting that changes in enzymatic efficiency are sufficient to explain the differential growth phenotypes (Fig. 2b) and the clade specificity observed (Fig 1a).

To determine if poorly active variants could be rescued by mutations in the co-evolving network, we conducted a random PCR-based mutagenesis of the entire S_G:E_E variant gene followed by two rounds of selection to isolate suppressors. We recovered an E184D mutation in 5 of 21 unique clones sequenced (Supplementary Fig. 10). This change of the metal-binding residue from E to D creates S_G:E_D from S_G:E_E, restored $k_{cat}$* and $K_M$* to wild-type levels (Fig. 3c), and resulted in a more rapid growth (Fig. 2b and Fig. 3a). Notably, a previous study that used random mutagenesis of another LHE, I-AniI, to select for increased activity identified the analogous E178D substitution, yet the co-evolutionary context of the substitution was not appreciated[20].

Collectively, our data overwhelmingly supports the existence of a network of co-evolving residues in LHEs, which can be predictably manipulated to modulate catalytic efficiency over an approximately 100-fold range. Efforts to re-engineer LHE for genome-editing applications have met with variable success[15,21-23], and we suggest this is due in part because such efforts inadvertently created a suboptimal complement of residues within the co-evolutionary network we have identified here. Our analyses predicted additional co-evolving networks in LHEs (Supplementary Table 1), experimental validation of which may be required to enhance success in re-engineering efforts. The consideration of co-evolving networks in other protein families could similarly add information to engineering of enzyme activity.

**METHODS SUMMARY**

**Sequence analyses.** An alignment of 178 non-redundant single-chain LHEs was constructed as previously described, and hand-edited to remove potential maturase sequences. Intra-molecular covariation was detected using LoCo[17], and visualized using MISTIC[24].

**Bacterial selection experiments.** Liquid competition and solid-media growth experiments utilized a two-plasmid bacterial selection system[18], with the I-LtrI or I-OnuI gene codon optimized for *Escherichia coli* and cloned into the pEndo expression plasmid. The corresponding target site cloned into the toxic plasmid (pTox). Selections were performed in *E. coli* NovaXGF' competent cells as described[18]. For liquid competition experiments, an I-LtrI library consisting of 1600 variants (synthesized by GenScript) was electroporated into NovaXGF' cells harboring pTox-Ltr, and harvested after 16 hrs growth under selective or non-selective conditions. For solid-media experiments, plasmids encoding individual I-LtrI or I-OnuI variants were electroporated into NovaXGF' competent cells, and plated on selective and non-selective media after either 1 hr or 4 hrs outgrowth to obtain a survival ratio.

**Illumina sequencing and bioinformatic analyses.** Bar-coded primers were used to amplify a region of I-LtrI gene encompassing positions 28, 29, 183 and 184, and paired-end sequencing was performed on an Illumina Mi-Seq at the London Regional Genomics Facility. Reads were parsed for the presence of D or E at positions 29 and 184, and then all possible quartets were identified using custom Perl and R scripts. The significance of the proportional abundance of each quartet in the selected versus non-selected condition was determined using an ANOVA-like differential analysis[19], modified for large datasets.

**Kinetic assays.** I-LtrI variants were purified using an N-terminal 6x-histidine tag that was removed post purification by Tev protease. Single-turnover conditions consisted of 10 nM supercoiled Ltr-target plasmid in reactions with I-LtrI ranging in concentration from 70 nM to 1391 nM. The initial reaction velocities were used to determine the pseudo-Michaelis-Menten parameters $k_{cat}$* and $K_M$*, as described [5,6].

**Supplementary Information** is linked to the online version of the paper at www.nature.com/nature.

**Acknowledgements** We thank Andrew Fernandes for discussions regarding statistical analyses. We thank the London Regional Genomics Facility and the Bi-Molecular Conformation and Interaction Facility for technical assistance with Illumina sequencing and differential scanning calorimetry analyses. R.J.D was supported by a Canada Graduate Scholarship from the National Science and Engineering Research Council of Canada. G.B.G is funded by a Discovery Grant from the National Science and Engineering Research Council of Canada, and D.R.E. is funded by an Operating Grant from the Canadian Institutes of Health Research (MOP 97780).

**Author Contributions** T.A.M. made the I-LtrI and I-OnuI constructs, created the mutant libraries, performed the bacterial selection experiments, and purified proteins for biophysical and kinetic analyses. R.J.D. performed phylogenetic and covariation analyses. S.T. performed bacterial selection experiments. G.B.G. and D.R.E. designed the study, performed bioinformatics analyses and wrote the paper with help from T.A.M. and R.J.D. T.A.M. and R.J.D. contributed equally to the paper. G.B.G. and D.R.E. are joint corresponding authors. All authors discussed the results and commented on the manuscript.

**FIGURE LEGENDS**

**Figure 1 | A co-evolving network in LHEs involving catalytic and non-catalytic residues**

A) A cladogram of single-chain LHEs generated from an unrooted maximum likelihood tree made by PhyML. The outer ring is coloured according to the identity of the metal-binding residues at positions 29 and 184, while the branches leading to the tips are coloured according to the identity of residues at positions 28 and 183, as indicated. Each clade is coloured by the residue 28_183 combination at the deepest branch, with no assumptions made regarding an ancestral state. The locations of the I-LtrI and I-OnuI endonucleases are denoted by triangles containing the first letter of the endonuclease.

B) Circos[24] plot illustrating covariation scores between residues of LHEs, mapped onto I-LtrI sequence. The outer ring shows I-LtrI residue number and identity for positions within the LHE alignment. The next ring shows amino acid conservation using a heat map (with red being most conserved and blue least conserved) and bar plot. The internal ring shows covariation scores using bar plots, with height of bar proportional to covariation score. Red lines connect positions with the highest covariation scores, black lines connect residues with intermediate scores, and grey lines connect positions with the lowest scores.

C) Covarying positions A28 (yellow), G183 (teal), E29 and E184 (red) mapped onto the structure of both I-LtrI (gold, PBD 3R7P) and I-OnuI (violet, PDB 3QQY) in the presence of DNA substrate. Magnesium cofactors (green) are shown as dotted spheres.

D) Heatmap of residue combinations in the LHE sequence alignment for positions 28, 29, 183 and 184 plotted on a log2 scale. Identities of the metal-binding residue at positions 29 and 184 are on the x-axis, and identities of residues at positions 28 and 183 are on the y-axis. Combinations of residues at positions 28 and 183 that were observed once in the LHE alignment (G_T, G_R, and G_D) are not plotted. ND, residue combination not detected in alignment.

**Figure 2 | Competition growth experiments reveal network preferences**

A) Schematic of the liquid competition growth experiment and sequencing strategy used to screen an I-LtrI mutant library.

B) Heatmap containing the log2 effect size[19] for quartet pairs with a false discovery rate of less than 0.05. The metal binding residue identities for 29 and 184 are found along the x-axis while the identity of residues 28 and 183 are found along the y-axis. Abundance scores for all 1600 possible quartets are found in Supplementary Table 3 and Supplementary Figure 2.

**Figure 3 | Effects of mutations in covarying network on catalysis and growth**

A) Heatmap depicting the log2 survival for individual I-LtrI (left) and I-OnuI (right) mutants in a bacterial two-plasmid selection. Identity of the metal-binding residues at positions 29 and 184 are on the x-axis, and identity of residues at positions 28 and 183 are on the y-axis. ND, not determined.

B) Boxplots of time (mins) for I-LtrI variants to reach mid-log growth ($A_{600}$ = 0.35) in selective media, with individual data points shown as dots. I-LtrI variants are indicated on the x-axis, with colors indicating variants that were analyzed kinetically in Figure 3c. Data for all quartets tested are shown in Supplementary Figure 5.

C) The $k_M$* (nM) and $k_{cat}$* (nM/min) values for select I-LtrI variants as determined by *in vitro* cleavage assays, plotted on a log10 scale. Three replicates for each variant are shown as open circles coloured according to quartet combination.

D) Plot of $k_M$* / $k_{cat}$* catalytic efficiency versus time to mid-log growth for I-LtrI variants on a log10 scale, coloured as in Figure 2c. Fit of the data to a linear regression model is shown by a black line, with the 95% confidence interval as a grey-shaded area. Plots for the relationship of $k_{cat}$* to mid-log growth, and $k_M$* to mid-log growth are shown in Supplementary Figure 9.

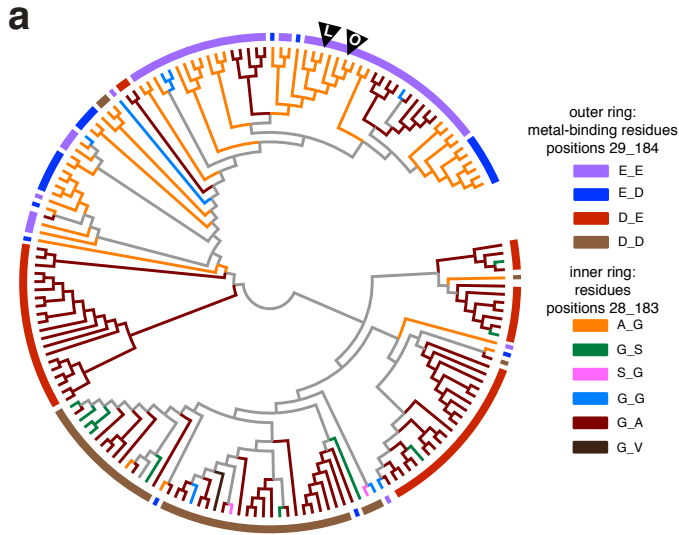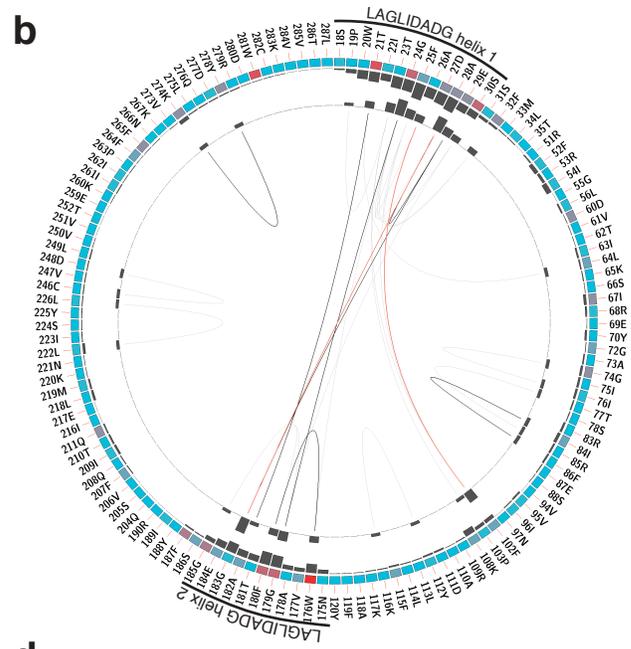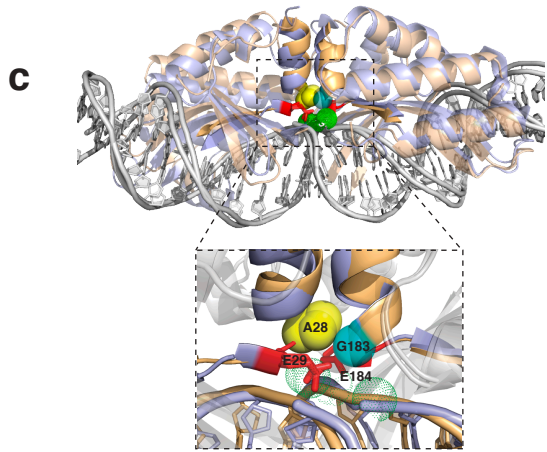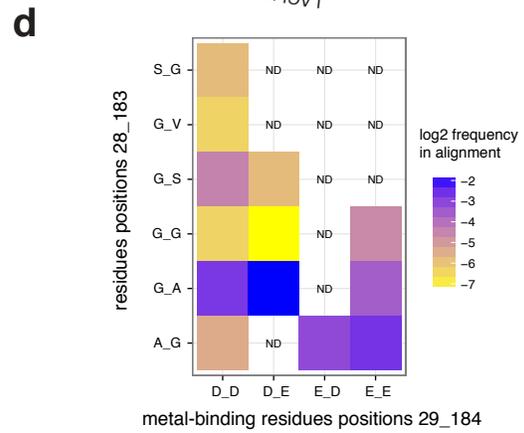

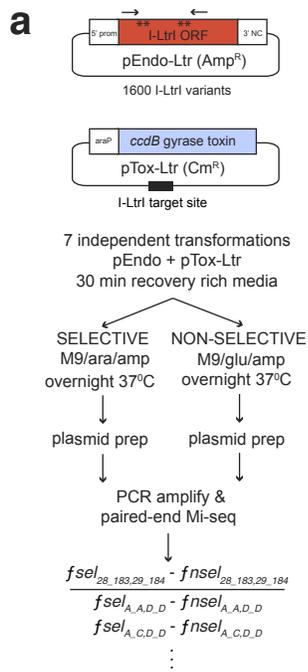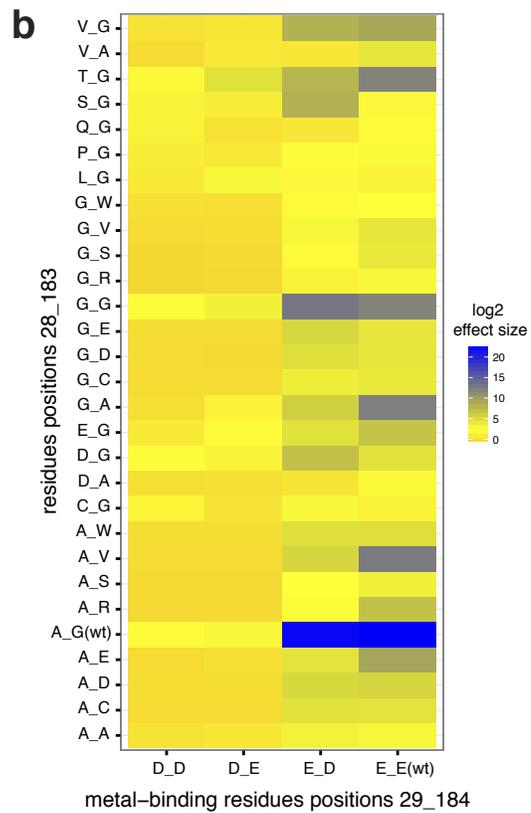

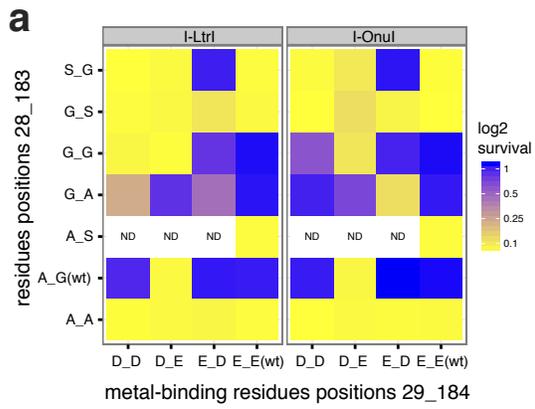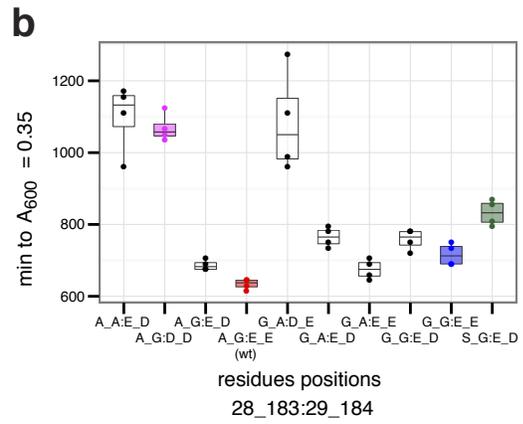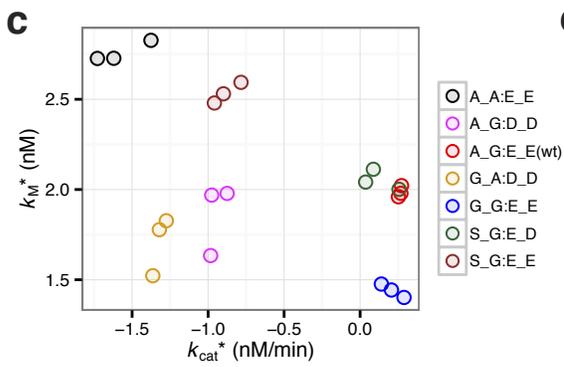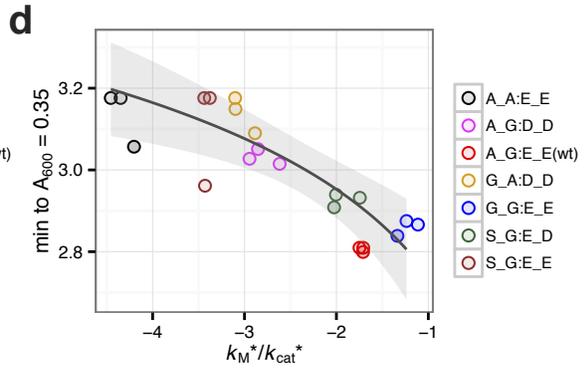